\documentclass[12pt,preprint]{aastex}
\usepackage{emulateapj5}

\newcommand {\simlt}{\lower.5ex\hbox{$\; \buildrel < \over \sim \;$}}
\newcommand {\simgt}{\lower.5ex\hbox{$\; \buildrel > \over \sim \;$}}

\begin{document}
\journalinfo{Accepted by Ap.J.Letters}
\title{Search for non-gaussian signals in the BOOMERanG maps: \\
pixel-space analysis}

\author{
G.~Polenta\altaffilmark{1}, P.A.R.~Ade\altaffilmark{2},
J.J.~Bock\altaffilmark{3}, J.R.~Bond\altaffilmark{4},
J.Borrill\altaffilmark{5}, A.~Boscaleri\altaffilmark{6},
C.R.~Contaldi\altaffilmark{4}, B.P.~Crill\altaffilmark{7}, P.~de
Bernardis\altaffilmark{1}, G.~De Gasperis\altaffilmark{8}, G.~De
Troia\altaffilmark{1}, K.~Ganga\altaffilmark{9},
M.~Giacometti\altaffilmark{1}, E.~Hivon\altaffilmark{9},
V.V.~Hristov\altaffilmark{7}, A.H.~Jaffe\altaffilmark{10},
A.E.~Lange\altaffilmark{7}, S.~Masi\altaffilmark{1},
P.D.~Mauskopf\altaffilmark{11}, A.~Melchiorri\altaffilmark{12},
T.~Montroy\altaffilmark{13}, P.~Natoli\altaffilmark{8},
C.B.~Netterfield\altaffilmark{14}, E.~Pascale\altaffilmark{6},
F.~Piacentini\altaffilmark{1}, D.~Pogosyan\altaffilmark{4},
S.~Prunet\altaffilmark{4}, G.~Romeo\altaffilmark{15},
J.E.~Ruhl\altaffilmark{13}, N.~Vittorio\altaffilmark{8},
A.~Zeppilli\altaffilmark{1} }

\affil{
$^{1}$ Dipartimento di Fisica, Universita' La Sapienza, Piazzale
 A. Moro 2, I-00185 Roma, Italy \\
$^{2}$ Queen Mary and Westfield College, London, UK \\
$^{3}$ Jet Propulsion Laboratory, Pasadena, CA, USA \\
$^{4}$ Canadian Institute for Theoretical Astrophysics,
        University of Toronto, Canada \\
$^{5}$ National Energy Research Scientific Computing Center,
        LBNL, Berkeley, CA, USA \\
$^{6}$ IROE-CNR, Firenze, Italy \\
$^{7}$ California Institute of Technology, Pasadena, CA, USA \\
$^{8}$ Dipartimento di Fisica, Universita' Tor Vergata,
        Roma, Italy \\
$^{9}$ IPAC, California Institute of Technology, Pasadena, CA, USA \\
$^{10}$ Astrophysics Group Blackett Laboratory, Imperial College, London,
UK \\
$^{11}$ Dept. of Physics and Astronomy, Cardiff University,
                Cardiff CF24 3YB, Wales, UK \\
$^{12}$ Nuclear and Astrophysics Laboratory, University of Oxford,
                Keble Road, Oxford, OX 3RH, UK\\
$^{13}$ Dept. of Physics, Univ. of California,
        Santa Barbara, CA, USA \\
$^{14}$ Depts. of Physics and Astronomy, University of Toronto, Canada \\
$^{15}$ Istituto Nazionale di Geofisica, Roma,~Italy \\
}

\begin{abstract}
We search the BOOMERanG maps of the anisotropy of the Cosmic
Microwave Background (CMB) for deviations from gaussianity. In
this paper we focus on analysis techniques in pixel-space, and
compute skewness, kurtosis and Minkowski functionals for the
BOOMERanG maps and for gaussian simulations of the CMB sky. We do
not find any significant deviation from gaussianity in the high
galactic latitude section of the 150 GHz map. We do find
deviations from gaussianity at lower latitudes and at 410 GHz, and
we ascribe them to Galactic dust contamination. Using non-gaussian
simulations of instrumental systematic effects, of foregrounds,
and of sample non-gaussian cosmological models, we set upper
limits to the non-gaussian component of the temperature field in
the BOOMERanG maps. For fluctuations distributed as a 1 DOF
$\chi^2$ mixed to the main gaussian component our upper limits are
in the few \% range.

\end{abstract}

\section{Introduction}\label{sec:intro}

In most inflationary scenarios the primordial density field is
expected to be gaussian \citep[]{Pee99}. In contrast, structure
formation scenarios based on topological defects \citep[]{Avel98}
 or less general inflationary
 models \citep[]{Linde97, Mart00, Cont99} predict a non-gaussian
density field. Thus, measurement of the statistical nature of the
CMB anisotropies can distinguish between these scenarios
\citep[]{Fisc85}.

The presence of noise in the measurements combined with the
limited coverage of the present observations can mask cosmological
non-gaussian features. Moreover, the presence of systematic
effects can produce subtle instrumental non-gaussian features in
intrinsically gaussian anisotropy maps.

Efforts to identify non-gaussianities in the CMB have been
extensively carried out for the COBE data \citep[]{Fer98, Band99,
Kom01, Brom99} and no significant detection of cosmological
non-gaussianity has been reported \citep{MagZZ}. However, the
sensitivity of COBE-DMR to the expected levels of cosmological
non-gaussianity from rare highly non-linear events like
topological defects \citep[]{Durr96} is not very high, due to the
large field of view
 which smears
out the effects of small scale features.
While the present observations on the power spectrum are inconsistent
with a structure formation scenario solely based on defects
\citep[]{Durr98}, a mixed
inflation+defects model is still compatible with the data
\citep[]{Bouc01}.

Furthermore, the low signal-to-noise ratio and coarse resolution
in the COBE maps also makes it difficult to detect primordial
non-gaussianity; the central-limit theorem states that the sum of
various instrumental effects will make the distribution tend to a
gaussian \citep[]{Novi00}.

Analyses to date of higher angular resolution experiments, like
QMASK \citep[]{Park01} and MAXIMA \citep[] {Sant01, Wu01}, show
full consistency with gaussianity.

In this paper we focus on the maps produced by the BOOMERanG experiment
\citep[]{deBe2000,Nett2001}.
Due to its wide sky and frequency
coverage, BOOMERanG is ideally suited to carry out an accurate analysis of
the possible systematic effects present in the detected signal.
We analyze them using a Monte-Carlo approach,
in order to set quantitative upper limits for the level of primordial
non-gaussian fluctuations present in the maps.

The data we use are presented in section ~\ref{sec:data}.  We use
five pixel-space estimators of departures from gaussianity: the
skewness and kurtosis of the CMB temperature distribution, and the
three Minkowski functionals: area, length and genus
\citep[]{Mink03}. In section ~\ref{sec:Analy} we apply these
estimators to the BOOMERanG maps and to gaussian Monte-Carlo
simulations of the CMB sky observed by BOOMERanG. In section
~\ref{sec:Syst} we analyze template maps of other non-gaussian
signals which could in principle be present in the BOOMERanG maps;
we study how sensitive the pixel-space techniques are in detecting
these effects, and we estimate how much these affect the
measurements of cosmological non-gaussian signals. In section
~\ref{sec:sour} we compare our measurements to the cosmological
expectations for a few sample scenarios.

The methods presented here are especially useful in the detection
of highly non-linear features (as expected in topological defects
theories or for Galactic foregrounds) \citep[]{Phil01}. One may
alternatively use a wavelets approach \citep[]{Hiv01}, or a
bispectrum approach \citep[]{Cont01}.

\section{The data}\label{sec:data}

We use the BOOMERanG data obtained in the Long Duration Balloon
(LDB) flight in 1998. For this paper we try to be as conservative
as possible, so we concentrate on the center of the map, 1.19\% of
the full sky  ($70^{\circ}<$RA$<105^{\circ}$, $-55^{\circ}<\delta
< -35^{\circ}$), where the sky coverage is approximately uniform
and the integration time per pixel is maximum. The galactic
latitude of this region spans from $-42^{\circ}$ to $-13^{\circ}$.
We also consider for comparison a lower latitude region with
similar size but extending to $b < -5^o$. We concentrate on the
best 150 GHz channel, B150A, for CMB signals, and on the best 410
GHz channel, B410B1, for monitoring the dust foreground. We use
only data collected in 1dps scans. Since pixel-space estimators
are sensitive to the size of the pixel, we have used 28', 14', and
7' pixels (HEALPix pixelisation \citep[]{Gors1998}). In fact, as
the pixel size is made smaller, smaller size structures become
visible, but the signal to noise ratio (SNR) per pixel decreases.
In the reference region, the noise per pixel at 150 GHz is of the
order of  30, 50, 90 $\mu K$ for 28', 14', and 7'pixels
respectively. For this reason it is important to investigate the
best tradeoff between angular resolution and SNR. The best fit
dipole anisotropy component is removed from the time ordered data
(TOD) before further processing. TOD are deconvolved from the
frequency response of the instrument and high-pass filtered with a
sharp filter, with a cut-on at 70 mHz. We use the improved
pointing solution of \citep[]{Nett2001}, featuring an equivalent
beam FWHM of 12.7' for the B150A channel. The time ordered data
obtained in this way are converted into a map using the fast
iterative generalized least squares map making of
\citep[]{Nato01}. The time-domain high-pass filtering procedure
removes most of the 1/f noise and most of the scan-synchronous
noise (SSN, see section 4), but also introduces an important
anisotropy in the large-scale content of the map: large scale
components of the sky temperature distribution are filtered out in
directions parallel to the scan, while being maintained in
directions orthogonal to the scan. This is not completely
recovered by the map making. Since the scan direction is not
constant during the flight, it is difficult to model analytically
the effect of such a filter. For this reason all the subsequent
analysis is performed numerically, using Monte-Carlo techniques.

\section{Analysis of the maps}\label{sec:Analy}

The normalized skewness $S_3$ and kurtosis $S_4$ are computed from
the $2^{nd}$, 3$^{rd}$ and 4$^{th}$ moments $\sigma^2=\sum_i(T_i-
\langle T \rangle)^2/(N-1)$, $\mu_3=\sum_i(T_i- \langle T \rangle
)^3/N$ and $\mu_4=\sum_i(T_i- \langle T \rangle )^4/N$ of the
distribution of the $T_i$: $S_3=\mu_3 / \sigma^3$, $S_4=\mu_4 /
\sigma^4 -3$. We find $S_{3,m}=-0.07$ and $S_{4,m} = -0.10$. These values
are plotted as vertical lines in fig.1. As a comparison, the 410
GHz channel in the same region produces $S_{3,m}=0.32$ and
$S_{4,m} = 1.01$. Due to the presence of sample variance, as well
as noise, correlations and anisotropic filtering, these values are
not expected to be exactly zero, even for a perfectly gaussian
sky. For this reason we compare the measured data to Monte-Carlo
simulations. These are based on gaussian maps of the CMB sky
extracted from a parent distribution with the best fit power
spectrum measured by BOOMERanG \citep[]{Nett2001}. These maps are
sampled along the BOOMERanG scans, simulated noise with the
statistical properties of the considered channel is added, and the
resulting TOD is filtered in the same way as the BOOMERanG data.
The SSN is not included in the simulation, but its contribution
is negligible (see section~\ref{sec:Syst}).
We then obtain simulated observed maps using the same algorithm we
have used for the actual BOOMERanG signals. From each simulated
map we compute $S_3$ and $S_4$, and plot their distribution in
fig.1. 

\begin{center}
\includegraphics[height=4.cm,width=8cm]{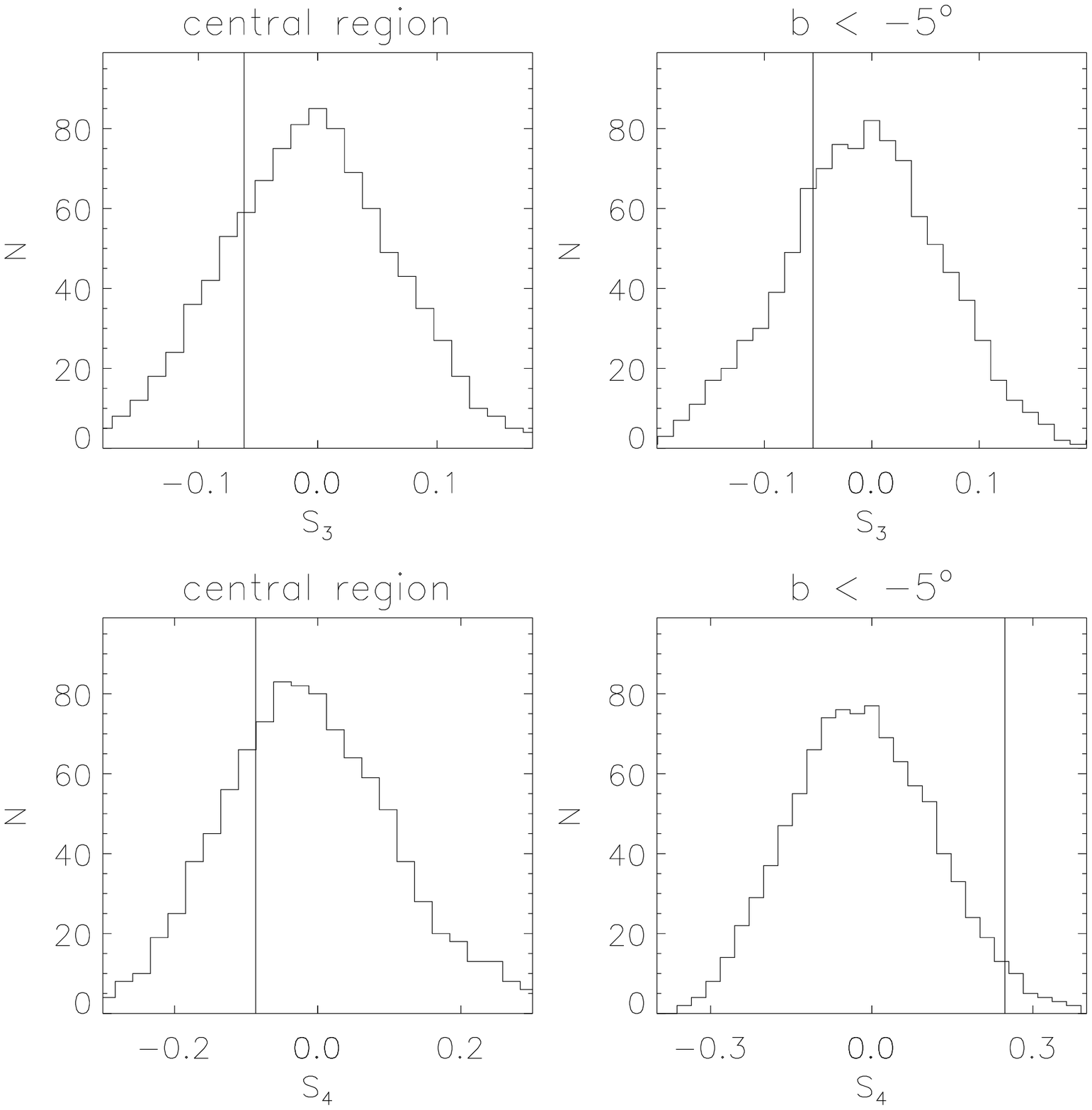}
\end{center}
\small FIG. 1. Measured values of the normalized skewness (top) and
kurtosis (bottom) in the nominal box (left) and in the low
latitudes box (right) at 150 GHz are shown as vertical lines. The
histograms are computed from detailed Monte-Carlo simulations of
the measurements. The simulations were obtained from a combination
of a gaussian sky with the best fit power spectrum measured in B98
and a simulation of detector noise. The different width of the
distributions is due to uneven coverage and noise per pixel.
\normalsize \vspace{0.2cm}

The visual comparison of the actual measurement to
the simulations suggests consistency with a gaussian sky at 150
GHz. From the distributions we compute the single-sided
probabilities $P(S_3>S_{3,m})=79\%$ and $P(S_4>S_{4,m})=73\%$ for
the 150 GHz channel, while for the 410 GHz channel
$P(S_3>S_{3,m})=2\%$ and $P(S_4>S_{4,m})=1\%$ . In the region at
lower Galactic latitudes $P(S_3>S_{3,m})=73\%$ and
$P(S_4>S_{4,m})=2\%$ at 150 GHz, and $P(S_3>S_{3,m})< 10^{-8}$ and
$P(S_4>S_{4,m})< 10^{-8}$ at 410 GHz. All the results above are
for a 28' pixel size. Similar results are obtained with 14' and 7'
pixels.

For a given threshold $\nu$ we compute the surface densities of
three Minkowski functionals $v_j$ following the contours ${\partial Q}$
of the excursion sets $Q\equiv Q(\nu)=\{ T_i: (T_i- \langle T
\rangle )/\sigma
> \nu \}$ \citep{Gott}:
\begin{equation} \label{eq:minkowski}
        v_0  =\frac{1}{A}\int_QdA ,~
        v_1  =\frac{1}{4A}\int_{\partial Q}ds ,~
        v_2  =\frac{1}{2\pi A}\int_
                {\partial Q}k ds
\end{equation}
where $dA$ and $ds$ are the differential elements of $Q$ and
$\partial Q$ respectively, and $k$ is the geodesic curvature of
$ds$.
The values of the $v_{j,m}$ measured from the 150 GHz map
are plotted as diamonds in fig.2. We show the results for 28'
pixels, since the CMB signal is larger than the noise in this
case. Again, even for a gaussian sky, the presence of noise,
correlations and anisotropic filtering does not allow us to
compare the measurements to the analytic expressions of $v_j$
which can be computed for gaussian isotropic distributed data. So
we compute the distributions of the $v_j$ from the simulated maps.
We plot the 95\% confidence band of the $v_j$ vs threshold $\nu$
in fig.2.

\begin{center}
\includegraphics[height=8.5cm,width=7.5cm]{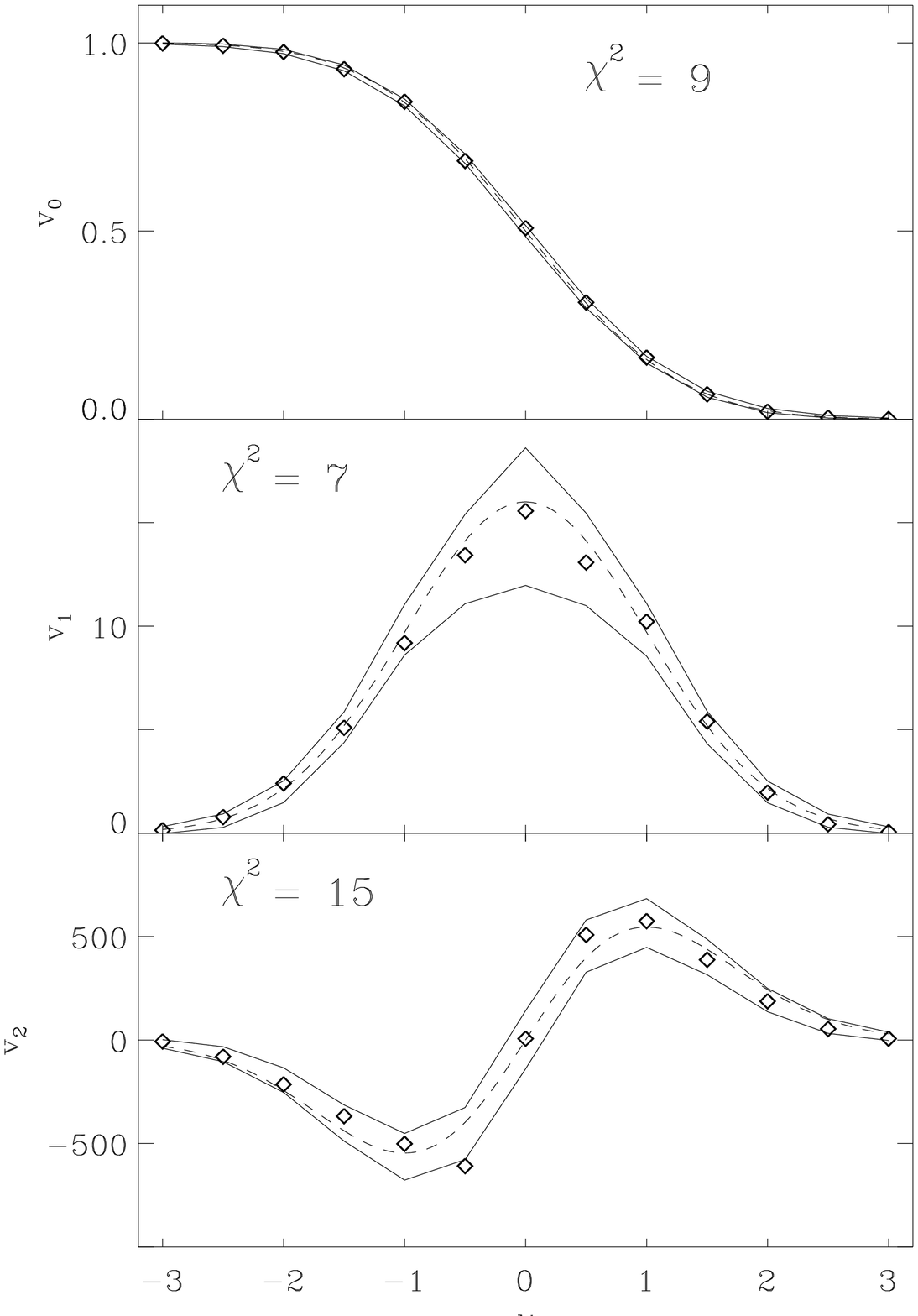}
\end{center}
\small FIG. 2. Measured values of the area (top), length (middle) and
genus (bottom) Minkowski functionals in the nominal box at 150 GHz
are plotted as diamonds vs the normalized threshold. The solid
lines represent the $95\%$ confidence band of the distributions of
the same quantities as computed from detailed Monte-Carlo
simulations of the measurements, obtained assuming a purely
gaussian sky with the best fit power spectrum measured in B98, and
5a realistic simulation of detector noise. Dashed lines are
5analytical predictions for the gaussian isotropic case. A 28'
pixelization has been used.
\normalsize \vspace{0.2cm}

 The consistency of the measured $v_{j,m}$ with the
gaussian sky hypothesis is evident. At this pixelization, the
$v_{j,m}$ are very similar to the ones computed from CMB only
simulations. At smaller pixelizations the effect of noise is to
increase $v_2$ and $v_3$ with respect to the CMB-only case, 
reducing the cosmological significance of the results.

To compare numerically the measured Minkowski functionals to the
simulations, we define 
\begin{eqnarray*}
 \chi^2_j & = & \sum_{\nu,\mu} \left[
v_j^{obs}(\nu) -\langle v_j^{sim}(\nu) \rangle \right]\cdot c^{-1}
\left[ v_j^{sim}(\nu), v_j^{sim}(\mu) \right]  \cdot \\
	& & \cdot \left[
 v_j^{obs}(\mu) -\langle v_j^{sim}(\mu) \rangle \right]
\end{eqnarray*}
 where $c
\left[ v_j^{sim}(\nu), v_j^{sim}(\mu) \right]$ is the covariance
matrix. From the measured data (with 28' pixels) we have
$\chi^2_0=9$, $\chi^2_1= 7$, $\chi^2_2= 15$, and $P(\chi^2
> \chi^2_0) = 76\%$, $P(\chi^2 > \chi^2_1) = 91\%$, $P(\chi^2 >
\chi^2_2) = 31\%$: again an excellent agreement with the gaussian
sky hypothesis. As a comparison, for the 410 GHz data we have
$\chi^2_0=23$, $\chi^2_1= 8$, $\chi^2_2=16$,
 $P(\chi^2 > \chi^2_0) = 0.1\%$,
$P(\chi^2 > \chi^2_1) = 84\%$, $P(\chi^2 > \chi^2_2) = 26\%$. In
the low Galactic latitudes box the Gaussian hypothesis is rejected
for all the functionals, for both the 150 GHz channel and the 410
GHz channel. Similar numerical results are obtained for 14' and 7'
pixels; we note, however, that for the finest resolution the
results are noise dominated rather than CMB dominated.

\section{Estimates of systematic effects}\label{sec:Syst}

Though we do not detect any evidence for non-gaussianity in the
maps, we set upper limits on possible sources of non-cosmological
non-gaussianity in order to assess the potential for further
analysis using more data. The raw BOOMERanG TOD are affected by a
large scale SSN, typically of the order of 20 $\mu K/^o$ in the
150 GHz channels and of 400 $\mu K/^o$ (CMB temperature units) in
the 410 GHz channels. The delay between the SSN and the azimuth
observed suggests an instrumental, thermal origin of the SSN. We
have subtracted the best fit CMB dipole from the 150 GHz TOD and
low passed the TOD to keep only the large scale SSN. We have then
built a map from this signal with the usual map making algorithm,
including the high pass filter, which strongly reduces the SSN
amplitude (see fig.3). Despite the relatively large signal in the
TOD, the anisotropies induced in the filtered map are of the order
of few $\mu K$ in the central box considered for this analysis.

 These are very small with respect to the detected
anisotropies, but evidently non-gaussian. We have used this map as
a template of a systematic effect. We built simulations of
contaminated data by adding this map to the gaussian simulations
of the CMB used in section ~\ref{sec:Analy}. Then we have applied
our gaussianity estimators to the simulated maps. The result is
that the distributions of the estimators are not affected by the
presence of SSN at the level we had in the 150 GHz channel
analyzed here. We compare the $\chi^2$ of the gaussian-only
simulations defined above to the $\chi^2$ of the contaminated
simulations, for different levels of the SSN map. We find that the
mean square amplitude of the SSN fluctuations should be of the
order of 35 to 70\% of the mean square CMB fluctuations (depending
on the estimator) before becoming detectable at 95\% C.L.. 
This means that the
SSN contribution would have to be at least 7 times larger than
that actually observed to affect the results reported in section
~\ref{sec:Analy}.

\begin{center}
\includegraphics[width=7cm,height=16cm]{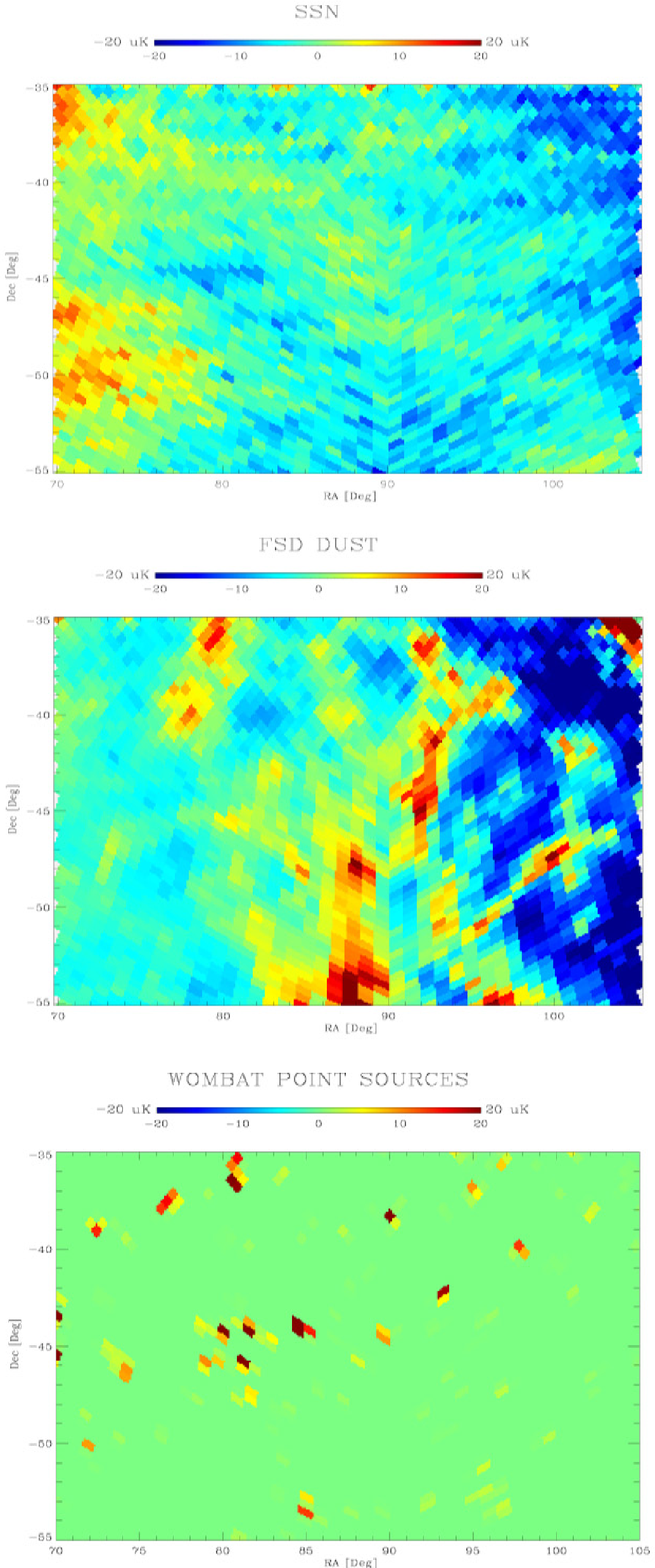}
\end{center}
\small FIG. 3. Estimated systematic effects at
 150 GHz in the observed sky region. The top panel shows the effect of
Scan
Synchronous Noise (SSN) when projected on the sky using the same
map-making procedure used for the actual data. The middle panel
represents the extrapolation of interstellar dust emission from
the FSD maps. The bottom panel shows the expected signal from
extragalactic point-sources extrapolated at 150 GHz using the
WOMBAT catalogue. 28' pixels have been used.
\normalsize \vspace{0.2cm}

The second source of a systematic effect we have considered is
thermal emission from interstellar dust. We have used the FSD maps
\citep[]{Fink99} filtered as in  \citep[]{Masi01} to produce a
template 150 GHz dust map. This is strongly non-gaussian, as
visible in fig.3, but its level is very small with respect to the
general level of CMB anisotropy. We have added this map to the
gaussian CMB simulations, and applied our gaussianity tests.
Again, the $\chi^2$ and the distributions of the contaminated
 simulations are very close to those of the gaussian-only simulations.
  Interstellar dust emission would have to be  3
times larger than the FSD  prediction  to contaminate the results
reported in section ~\ref{sec:Analy} at 95\% C.L..

The third  source of a systematic effect we have considered is
emission from extragalactic sources present in the observed
region. We have used the WOMBAT catalogue \citep[]{WO01} to
extrapolate to 150 GHz the flux of known radio sources in the
surveyed area. From the catalogue obtained in this way we create a
map convolving the point-source flux with the angular response of
our telescope. The map (fig.3)  --- which is evidently
non-gaussian  ---  is added to the gaussian CMB simulations and
the analysis with our estimators is carried out. Yet again, the
$\chi^2$ and the distributions of the contaminated simulations are
very close to  those of the gaussian-only simulations. Fluxes from
extragalactic sources would have to be 6 times larger than the
WOMBAT  predictions to contaminate the results reported in section
~\ref{sec:Analy} at 95\% C.L..

\section{Cosmological Sources of non gaussianity}\label{sec:sour}

There are many models of non-gaussian CMB fluctuations. We have
considered a few which can be easily compared to our results. As a
first concrete example, we have used Linde's model
\citep[]{Linde97}. In this case CMB fluctuations are distributed
as a $\chi^2$ with 1 DOF. This hypothesis is strongly rejected by
all of our estimators. In other models there is a mixture of
gaussian and non-gaussian CMB anisotropies (e.g.
\citep[]{Bouc01}). For example, Lyth and Wands
(\citep[]{Lyth2k1}), consider a mixture of 1 DOF $\chi^2$
distributed fluctuations with Gaussian CMB anisotropies. From
Monte Carlo simulations we find that the 95$\%$ confidence upper
limits to the rms amplitude of the non-Gaussian signal are 5$\%$,
6$\%$ and 6$\%$ of the Gaussian one (using Area, Length and Genus
Minkowski estimators respectively, with 28' pixels); the upper
limits become 2$\%$, 5$\%$ and 8$\%$ respectively, with 7' pixels.
The upper limit we get is not very sensitive to the pixelization because 
the signal to noise ratio increases by increasing the pixel size, 
but the angular resolution decreases, and there are less spots on which 
the functionals can be built. The two effects roughly compensate each other.
                                  
We have also simulated CMB maps with a specified (equal-$\ell$
reduced) bispectrum $B_\ell = A C_\ell^{3/2}$ (see
\citep[]{Con01}). For $C_\ell$ we used the B98 best fit power
spectrum. For the non-Gaussianity estimators considered in this
paper, we detected no deviation from Gaussianity significant at
the 95$\%$ CL over the range of amplitudes $A$ from 0 to 10.

\section{Conclusions}\label{sec:Disc}

The pixel-based techniques discussed here and applied to the
central region observed by BOOMERanG at 150 GHz confirm the
Gaussianity of the detected CMB fluctuations, and exclude with
high confidence $\chi^2$ distributed CMB temperature fluctuations.
These techniques also detect non-Gaussian fluctuations due to
interstellar dust at high frequencies and at lower latitudes.
Subdominant non-Gaussian fluctuations mixed to the Gaussian ones
are strongly excluded for some models, while are not detected at
all in other cases.

\acknowledgments

The BOOMERanG project has been supported by Programma Nazionale di
Ricerche in Antartide, Universit\'a di Roma ``La Sapienza'', and
Agenzia Spaziale Italiana in Italy, by NASA and by NSF OPP in the
U.S., by PPARC in the UK, and by the CIAR and NSERC in Canada. We
acknowledge the use of the HEALPix package. We thank the referee
for useful suggestions.

\end{document}